# Lifshitz Formula Using Box Renormalization

Maryam Miralaei and Reza Moazzemi

Department of Physics, University of Qom, Ghadir Blvd., Qom 371614-611, I.R. Iran

October 3, 2018

## Abstract

In this paper the Lifshitz formula for the Casimir energy between two dielectrics in zero temperature is derived using box renormalization. Although there are several derivations for the force in this case in the literature, including Lifshitz's own proof, so far there has been no unambiguous and rigorous derivation for energy that we are aware of. Since the energy becomes important in some cases, e.g. calculation of entropy or heat capacity, using the correct and precise definition of the Casimir energy, for the first time, we remove all of the infinities systematically without any ambiguity. This proof also shows the strength and accuracy of the box renormalization scheme.

# فرمول لیفشیتز با استفاده از بازبهنجارش جعبه

مریم میرعلائی و رضا معظمی*

گروه فیزیک، دانشکده علوم، دانشگاه قم، قم، ایران

## چکیده

در این مقاله فرمول لیفشیتز برای انرژی کازیمیر بین دو دی‌الکتریک در دمای صفر مطلق با استفاده از روش بازبهنجارش جعبه به دست می آید. هر چند که برای محاسبه نیروی کازمیر در این مورد در منابع مختلف اثبات های متعددی ارائه شده است از جمله اثبات خود لیفشیتز، اما ما تاکنون اثبات بدون ابهام و بسیار دقیق برای انرژی ندیده‌ایم. از آنجا که خود انرژی در بعضی موارد، مانند محاسبه آنتروپی یا ظرفیت گرمایی مهم می شود، ما اینجا از تعریف صحیح و دقیق انرژی کازیمیر استفاده می کنیم و، برای اولین بار، کلیه بینهایت‌ها را به صورت سیستماتیک و نه دستی، بدون ابهام حذف خواهیم نمود. این اثبات همچنین قدرت و دقت روش بازبهنجارش جعبه را نشان می دهد.

**کلیدواژگان:** انرژی کازیمیر، فرمول لیفشیتز، خلأ کوانتومی، میدان الکترو مغناطیسی، بازبهنجارش جعبه.

## مقدمه

اثر کازیمیر در سال ۱۹۴۸ میلادی توسط هندریک کازیمیر،[1] فیزیکدان هلندی، معرفی شد[۱]. کازیمیر وجود نیروی جاذبه‌ای را بین دو صفحه‌ی کاملاً رسانا، بدون بارخالص الکتریکی، موازی و صاف که در دمای صفر مطلق و در فاصله‌ی میکرومتری $a$ در خلأ کوانتومی قرار داده شده‌اند، پیش‌بینی کرد.

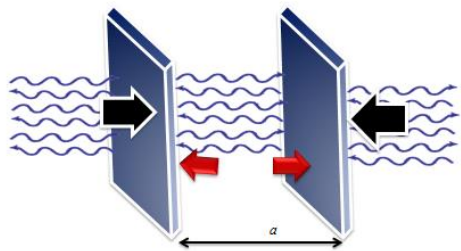

شکل ۱: نیروی کازیمیر بین دو صفحه‌ی رسانا.

طبق محاسبه‌ی وی مقدار این نیرو در واحد سطح برابر با

$$F_{Cas} = -\frac{\hbar c \pi^2}{240}\frac{1}{a^4},  \qquad ۱$$

می‌باشد، که در آن $c$ و $\hbar$ به ترتیب سرعت نور و ثابت پلانک است. همانطور که مشاهده می‌شود این نیرو برای این پیکر بندی متناسب با عکس توان چهارم فاصله‌ی بین دو صفحه می‌باشد. این پدیده یک اثر کاملاً کوانتومی است که به دلیل اُفت و خیزهای انرژی خلأ کوانتومی به وجود می‌آید. به بیان دیگر انرژی خلأ در حضور و عدم حضور صفحات یکسان نمی‌باشد، طوری که اختلاف این دو انرژی به فاصله‌ی بین صفحات بستگی دارد. مشتق این اختلاف نسبت به فاصله‌ی $a$ به ما نیرو می‌دهد. در واقع این اختلاف

---

انرژی، برابر با کاری است که باید صورت پذیرد تا دو صفحه از بی‌نهایت آورده به فاصله‌ی $a$ از یکدیگر قرار داده شوند. این نیرو به علت قدرت جاذبه‌ی بالا در فواصل میکرومتری و زیر میکرومتری، در طراحی سیستم‌های میکرو مکانیکی قابل توجه خواهد بود[۲]. اوُلین آزمایش در سال ۱۹۵۸میلادی توسط مارکوس اسپارنای[1] به ثبت رسید[۳]. هرچند که وی بر مشکلات آزمایشگاهی فایق آمد اما نتیجه‌ی آزمایش او همراه خطای ۱۰۰درصدی بود، با این وجود نظریه‌ی کازیمیر را رد نکرد. بعد از ایشان آزمایشات متعددی برای اثبات اثر کازیمیر انجام شد اما به دلیل خطای بالا مورد اهمیت قرار نگرفتند. در سال ۱۹۹۶میلادی عمر محی الدین[2] موفق به آزمایشی شد که ۹۹ درصد موافق نظریه‌ی پیش بینی شده‌ی کازیمیر بود[٤].

البته باید توجه داشت که رسانای کامل در طبیعت وجود ندارد. به همین علّت لیفشیتز[3]در سال ۱۹۵۶ میلادی نیروی کازیمیر را برای دو صفحه‌ی دی‌الکتریک ضخیم به دست آورد[٥]. البته خود او جزئیات محاسبه را در مقاله نیاورده و تنها روش به دست آوردن فرمول را بیان کرده است.

بسیاری از نویسندگان سعی کرده‌اند که فرمول وی را به روش‌های مختلف به دست آورند، به عنوان نمونه منابع[۱۳-٦] را ببینید. اما در هیچ یک از این منابع، ما روش سیستماتیک و خودسازگاری جهت استخراج این فرمول نیافتیم. به عنوان مثال در بعضی منابع نویسندگان از تعریف انرژی کازیمیر به صورت

$$E_{Cas}(a) = E_0(a) - \lim_{a \to \infty} E_0(a) \qquad ۲$$

استفاده نموده‌اند[۲٦-۸]، که اگر منظور از $E_0(a)$ انرژی بین صفحات باشد، این تعریف با تعریف اصلی آن که تفاضل انرژی نقطه‌ی صفر کل خلأ با حضور و عدم حضور صفحات است سازگار نمی‌باشد. در معادله‌ی (۲) اگر منظور از $E_0(a)$ انرژی بین صفحات باشد، انرژی خلأ مربوط به بین صفحات از انرژی کل خلأ بدون صفحه کسر می‌شود که این معنی فیزیکی خاصی ندارد. لذا طبیعی است که انتظار داشته باشیم جواب ایشان بی‌نهایت باشد. در بعضی منابع اثر این اشتباه در تعریف با سایر اشتباهات محاسباتی از بین رفته و در نهایت به فرمول اصلی رسیده‌اند. در برخی منابع دیگر برای برطرف کردن بی‌نهایت‌های مسأله روش‌های پیچیده‌ی دیگری مانند روش پراکندگی مُدها[۷,٦] ارائه نموده‌اند. این روش‌ها علاوه بر اختلاف و تفاوتی که با روش معمولی و سرراست کم کردن انرژی‌ها دارند خود نیز دچار ابهام هستند که در ادامه توضیح خواهیم داد. در محاسبه‌ی انرژی کازیمیر همواره با تفاضل دو انرژی بی‌نهایت برخورد داریم، که طبیعتاً جواب آن از نظر ریاضی مبهم می‌باشد:

$$E_{Cas} = \frac{\hbar}{2}\sum \omega - \frac{\hbar}{2}\sum \omega_0 \qquad ۳$$

که در آن ($\omega_0$ها) $\omega$ها مُدهای میدان کوانتومی مورد نظر در (عدم) حضور صفحات می‌باشند. برای استخراج جواب فیزیکی از تفاضل دو بی‌نهایت باید بسیار محتاط و دقیق عمل نمود. باید دقت کرد هنگامی که صفحات وجود دارند ما از شرایط مرزی غیربدیهی (دیریکله،

---

نیومن و ...) برای به دست آوردن انرژی خلأ، استفاده می‌کنیم، اما برای خلأ بدون صفحه معمولاً از شرایط مرزی دوره‌ای استفاده می‌شود. این اختلاف خود ممکن است باعث ابهام در جواب نهایی مسأله شود. بعنوان مثال فرمول مرسومی که در مقالات از آن استفاده می‌شود به شکل

$$E_{Cas}(a) = E_0(a) - \lim_{L \to \infty} \frac{a}{L} E_0(L)$$

است، که هرچند در بسیاری از موارد این فرمول به جواب صحیح می‌انجامد اما در این مسأله به دلایلی که ذکر شد هنوز مبهم باقی می‌ماند.

در این مقاله ما قصد داریم با یک روش سیستماتیک و خودسازگار از تعریف اصلی و دقیق انرژی کازیمیر فرمول لیفشیتز را استخراج نماییم. روش ما حذف بی‌نهایت ها با بازبهنجارش جعبه می‌باشد[۲۹-۳۲] ایده‌ی اساسی ما اینجا این است که با قرار دادن کل سیستم در یک جعبه، در هر دو حالت از یک نوع شرایط مرزی استفاده شود و انتظار داریم که در این صورت بی‌نهایت‌ها به صورت سیستماتیک و نه دستی از مسأله حذف شوند. خواهیم دید که بهنجارش جعبه این کار را برای ما انجام می‌دهد. ما ابتدا در بخش بعد انرژی کازمیر را با روش بازبهنجارش جعبه توضیح می‌دهیم، و در ادامه خواهیم دید که این انرژی به جمع مُدها وابسته است. لذا با استفاده از توابع مولد مُدها که در پیوست به دست می‌آید به اثبات رابطه انرژی لیفشیتز با استفاده از بازبهنجارش جعبه خواهیم پرداخت. لازم به ذکر است هر چند فرمول لیفشیتز هم برای دمای صفر و هم برای دمای غیر صفر بیان شده، مساله ای که ما در اینجا به بررسی آن می‌پردازیم در دمای صفر مطلق است، لذا از نظریه میدان کوانتومی استاندارد در دمای صفر استفاده می‌شود و کمینه انرژی کل محاسبه خواهد شد.

## بررسی فرمول لیفشیتز بین دو صفحه‌ی دی الکتریک

در این بخش برای به دست آوردن فرمول لیفشیتز ابتدا روش بازبهنجارش جعبه را بیان می‌کنیم سپس برای به دست آوردن مُدها با استفاده از حل معادلات ماکسول شرایط مرزی را اعمال می‌کنیم.

### بازبهنجارش جعبه

همانطور که در مقدمه اشاره شد، برای کم کردن صحیح بی‌نهایت‌ها از یکدیگر (بهنجارش) در دو حالت حضور و عدم حضور صفحات ما کل دستگاه را در هر حالت در یک جعبه قرار می‌دهیم و انرژی کل این دو پیکربندی را از یکدیگر کم می‌کنیم. در حالت اوّل دو صفحه‌ی موازی در فاصله‌ی $a$ از یکدیگر، در نقاط $\pm\frac{a}{2}$، درون جعبه‌ای با ابعاد $L \times S$، در خلأ قرار داده شده‌اند (شکل ۲ (بالا)). $S$ مساحت دیواره‌های جعبه (یا صفحات) و $L$ فاصله‌ی آنها از هم می‌باشد. در پیکر بندی دوّم دو صفحه را در فاصله $b$ از یکدیگر درون جعبه مشابه‌ای قرار می‌دهیم که البتّه در نهایت برای شرط عدم حضور صفحات $b$ را به بی‌نهایت میل خواهیم داد (شکل ۲ (پایین)). در این بهنجارش انرژی کازیمیر به صورت زیر تعریف می‌شود[۲۹-۳۲]:

$$E_{\text{Cas}}(a) = \lim_{b \to \infty} \lim_{L \to \infty} \left[ E(a,L) - E(b,L) \right] \quad \text{٤}$$

که در آن $E(a,L)$ و $E(b,L)$ به ترتیب انرژی کل پیکربندی‌های اوّل و دوّم می‌باشند، که اگر $E(a)$ را مقدار انتظاری انرژی خلأ بین دو صفحه به فاصله‌ی $a$ در نظر بگیریم به صورت زیر نوشته می‌شوند:

$$E(a,L) = E(a) + 2E\left(\frac{L-a}{2}\right)$$

$$E(b,L) = E(b) + 2E\left(\frac{L-b}{2}\right).$$

۵

باید دقت شود که در این مساله صفحات را به اندازه کافی ضخیم در نظر می‌گیریم که میدانها به خوبی در آن افت کرده و میدانهای بیرون نتوانند به داخل نفوذ کنند، همچنین در معادله (۵) چون در نهایت $L$ را به بی‌نهایت میل می‌دهیم ضخامت صفحه‌ها را در معادله وارد نکرده‌ایم (یعنی مثلا $L+d$ ننوشتیم). با توجه به اینکه هم جنس بودن صفحات منجر به این می‌شود که شرایط مرزی یکسانی روی آنها اعمال شود، مزیّت این روش در این است که اگر صفحات و دیواره‌های جعبه را مشابه در نظر بگیریم، در هر دو پیکربندی انرژی‌ها با روش‌های شبیه هم به دست می‌آید و انتظار ما این است که جواب نهایی بعد از تفاضل، متناهی و فاقد بی‌نهایت‌های اضافی باشد.

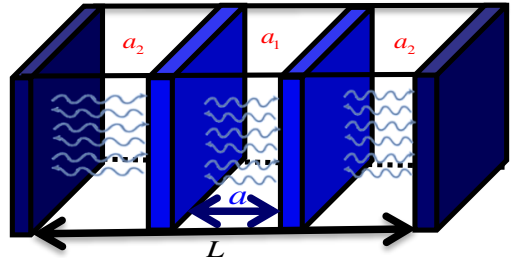

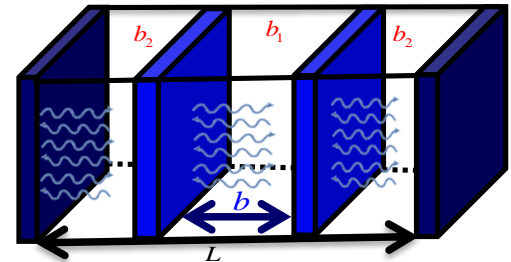

شکل ۲: در شکل بالا (پایین) دو صفحه به فاصله‌ی $a$ ($b$) از یکدیگر درون جعبه با ابعاد $L\times S$ قرار دارند. برای انرژی کازیمیر لازم است تفاضل انرژی کل این دو پیکربندی را به دست آوریم سپس $L$ و $b$ را به ترتیب به بی‌نهایت میل دهیم. لازم به ذکر است برای اینکه در هنگام حدگیری هر دو صفحه به بی‌نهایت برده شوند ما صفحات را در $\pm\dfrac{b}{2}$

و $\pm\dfrac{a}{2}$ و دیواره‌های جعبه را در $\pm\dfrac{L}{2}$ قرار داده‌ایم. همچنین باید دقت شود ابتدا حد $L\to\infty$ و سپس حد $b\to\infty$ گرفته خواهد شد. نیروی کازیمیر با مشتق‌گیری از رابطه‌ی (۴) به صورت زیر به دست می‌آید:

$$F_{\text{Cas}}(a)=-\dfrac{dE_{\text{Cas}}(a)}{da}. \qquad ۶$$

**مقدار انتظاری انرژی خلأ و توابع مولد مُدها**

همانطور که ذکر شد برای به دست آوردن نیروی کازیمیر از مشتق انرژی یعنی همان مؤلفه‌ی $T^{00}$ تانسور انرژی- تکانه استفاده می‌کنیم (روش دیگر، محاسبه‌ی مستقیم فشار با استفاده از مؤلفه‌ی $T^{zz}$ می‌باشد)، یعنی

$$\begin{aligned}E_0=\langle H\rangle_0&=\langle 0|T^{00}|0\rangle\\&=\langle 0|\int d^3x\,\mathscr{T}^{00}|0\rangle.\end{aligned} \qquad ۷$$

که در آن

$$\mathscr{T}^{\mu\upsilon}=\dfrac{\partial\mathscr{L}}{\partial\partial_\mu A_\sigma}\partial^\upsilon A_\sigma-g^{\mu\upsilon}\mathscr{L} \qquad ۸$$

تانسور چگالی انرژی تکانه می‌باشد و لاگرانژین میدان الکترومغناطیس به صورت

$$\mathscr{L}=-\dfrac{1}{4}F_{\mu\upsilon}F^{\mu\upsilon} \qquad ۹$$

است، که در آن $F_{\mu\upsilon}$ تانسور اندازه میدان می‌باشد که به شکل زیر است

$$F^{\mu\upsilon}=\partial^\mu A^\upsilon-\partial^\upsilon A^\mu. \qquad ۱۰$$

میدان کوانتومی برداری $A_\mu$ همان میدان یک فوتون، میدان الکترومغناطیس، می‌باشد. شکل این میدان بعد از کوانتش دوّم به صورت

$$A_\mu(x) = \sum_{\mathbf{k}} \sum_{s=1,2} \left(\frac{\hbar c^2}{2V\omega_{\mathbf{k}}^s}\right)^{\frac{1}{2}} \qquad 11$$

$$\times \varepsilon_s^\mu(\mathbf{k}) \left[a_s(\mathbf{k})e^{-ikx} + a_s^\dagger(\mathbf{k})e^{ikx}\right]$$

است، که در آن $s$ شمارنده دو نوع قطبش فوتون است و $a_s^\dagger(\mathbf{k})$ و $a_s(\mathbf{k})$ به ترتیب عملگرهای خلق و فنای یک فوتون با تکانه‌ی $\mathbf{k}$، انرژی $\hbar\omega_{\mathbf{k}}^s$ و قطبش $\varepsilon_s^\mu(\mathbf{k})$ می‌باشند. با استفاده از این میدان و معادله‌ی (8) خواهیم داشت

$$T^{00} = \sum_{s\mathbf{k}} \hbar\omega_{\mathbf{k}}^s \left[a_s^\dagger(\mathbf{k})a_s(\mathbf{k}) + \frac{1}{2}\right]. \qquad 12$$

در نتیجه مقدار انتظاری انرژی خلأ، معادله‌ی(7)، به صورت

$$E_0 = \frac{\hbar}{2} \sum_{s\mathbf{k}} \omega_{\mathbf{k}}^s \qquad 13$$

خواهد شد. در رابطه‌ی فوق برای هر یک از راستاهایی که شرط مرزی غیر بدیهی اعمال نشود می‌توان جمع را به انتگرال تبدیل نمود ($\frac{1}{L}\sum \to \int \frac{dk}{2\pi}$). مثلاً اگر تنها در راستای $z$ شرط مرزی غیر بدیهی داشته باشیم و در دو راستای عمود دیگر شرط مرزی بدیهی باشد، مانند دو صفحه‌ی موازی عمود بر محور $z$ به فاصله‌ی $a$، $E_0$ به صورت زیر خواهد بود:

$$E(a) \equiv E_0 = \frac{\hbar}{2} \int \frac{d^2\mathbf{k}_\perp}{(2\pi)^2} \sum_n \omega_{\mathbf{k},n} S. \qquad 14$$

برای به دست آوردن بسامدهای $\omega$ در این مسأله، فرض می‌کنیم که دو صفحه‌ی دی الکتریک با ثابت گذردهی نسبی $\varepsilon(\omega)$ به فاصله‌ی $a$ از یکدیگر قرار گرفته‌اند. با حل معادلات ماکسول و اعمال شرایط مرزی روی صفحات، توابع مولد مُدها را به دست خواهیم آورد. جزئیات این کار در پیوست آمده است. برای دو نوع قطبش TM و TE، مغناطیسی عرضی و الکتریکی عرضی، توابع مولد مُدها به صورت زیر به دست می‌آیند:

$$\Delta^{TM}(\omega, \kappa_\perp, a) = e^{-\kappa a}\left[(\varepsilon q + \kappa)^2 e^{qa}\right.$$
$$\left. - (\varepsilon q - \kappa)^2 e^{-qa}\right] \qquad 15$$

$$\Delta^{TE}(\omega, \kappa_\perp, a) = e^{-\kappa a}\left[(q + \kappa)^2 e^{qa}\right.$$
$$\left. - (q - \kappa)^2 e^{-qa}\right], \qquad 16$$

که در آن کمیت‌های $\kappa$ و $q$ از رابطه‌ی (پ.11) و (پ.12) به صورت

$$\kappa_\perp = (\kappa_x, \kappa_y) \qquad q^2 = \kappa_\perp^2 - \frac{\omega^2}{c^2},$$

$$\kappa^2 = \kappa_\perp^2 - \varepsilon(\omega)\frac{\omega^2}{c^2}, \qquad 17$$

می‌باشند. ریشه‌های $\Delta^{TM,TE}$ همان مُدهای مجاز می‌باشند. لذا انرژی خلأ میدان الکترومغناطیسی در حضور صفحات دی‌الکتریک واقع در فاصله‌ی $\pm\frac{a}{2}$ در دمای صفر به صورت زیر می‌شود

$$E(a) = \frac{\hbar}{4\pi} \int_0^\infty \kappa_\perp d\kappa_\perp$$
$$\times \sum_n \left(\omega_{\kappa_\perp, n}^{TM} + \omega_{\kappa_\perp, n}^{TE}\right) S. \qquad 18$$

برای محاسبه‌ی انرژی می‌توان طبق اصل آرگومان[34] به یک انتگرال پربندی در فضای مختلط $\omega$ تبدیل نمود. اصل آرگومان برای یک تابع $g(z)$ که در ناحیه‌ی پربندی تحلیلی می‌باشد، به شکل زیر نوشته می‌شود

$$\sum_n g(a_n) - \sum_m g(b_m) = \frac{1}{2\pi i}$$
$$\times \oint_C g(z) d\ln\Delta(z) \qquad 20$$

که در آن $a_n$ ها و $b_m$ ها به ترتیب ریشه‌ها و قطب‌های تابع $\Delta(z)$ می‌باشند و پربند بسته $C$ کلیه‌ی ریشه‌ها و قطب‌ها را در بر می‌گیرد. در نتیجه معادله‌ی (۱۸)، به شکل زیر خواهد شد:

$$E(a) = \frac{\hbar}{4\pi} \int_0^\infty \kappa_\perp d\kappa_\perp$$
$$\times \frac{1}{2\pi i} \oint_C \omega d\ln\Delta^{TM,TE}(\omega,\kappa_\perp) S \qquad 19$$

در رابطه‌ی فوق پربند بسته‌ی $C$ را مانند شکل ۳ در نظر می‌گیریم. این پربند شامل یک نیم دایره به شعاع بی‌نهایت در نیمه‌ی راست صفحه‌ی مختلط $\omega$، $C_+$، و یک خط روی محور موهومی $(-i\infty, +i\infty)$ به صورت پادساعتگرد می‌باشد. حال با جایگذاری رابطه‌ی (۲۰) در معادله‌ی (۴) (مثلاً برای مُد TM) خواهیم داشت:

$$E_{Cas}^{TM}(a) = \lim_{b\to\infty}\lim_{L\to\infty}\frac{\hbar S}{8\pi^2 i}\int_0^\infty \kappa_\perp d\kappa_\perp \qquad 21$$
$$\times \oint_C \omega d\ln\left[\frac{e^{-2\kappa(\frac{L-a}{2})}\left((\varepsilon q+\kappa)^2 e^{q(\frac{L-a}{2})} - (\varepsilon q-\kappa)^2 e^{-q(\frac{L-a}{2})}\right)^2 e^{-ka}\left((\varepsilon q+\kappa)^2 e^{qa} - (\varepsilon q-\kappa)^2 e^{-qa}\right)}{e^{-2\kappa(\frac{L-b}{2})}\left((\varepsilon q+\kappa)^2 e^{q(\frac{L-b}{2})} - (\varepsilon q-\kappa)^2 e^{-q(\frac{L-b}{2})}\right)^2 e^{-kb}\left((\varepsilon q+\kappa)^2 e^{qb} - (\varepsilon q-\kappa)^2 e^{-qb}\right)}\right].$$

بعد از گرفتن حد $L \to \infty$ عبارت فوق به صورت زیر ساده خواهد شد

$$E_{Cas}^{TM}(a) = \lim_{b\to\infty}\frac{\hbar S}{8\pi^2 i}\int_0^\infty \kappa_\perp d\kappa_\perp \oint_C \omega d\ln\left[\frac{(\varepsilon q+\kappa)^2 e^{qa}\left((\varepsilon q+\kappa)^2 e^{qa} - (\varepsilon q-\kappa)^2 e^{-qa}\right)}{(\varepsilon q+\kappa)^2 e^{qb}\left((\varepsilon q+\kappa)^2 e^{qb} - (\varepsilon q-\kappa)^2 e^{-qb}\right)}\right]$$

$$= \frac{\hbar}{8\pi^2 i}\int_0^\infty \kappa_\perp d\kappa_\perp \oint_C \omega d\ln\left[\frac{e^{-qa}\left((\varepsilon q+\kappa)^2 e^{qa} - (\varepsilon q-\kappa)^2 e^{-qa}\right)}{(\varepsilon q+\kappa)^2}\right]$$

$$= \frac{\hbar}{8\pi^2 i}\int_0^\infty \kappa_\perp d\kappa_\perp \oint_C \omega d\ln\left(1 - \left(\frac{\varepsilon q-\kappa}{\varepsilon q+\kappa}\right)^2 e^{-2qa}\right), \qquad 22$$

به این ترتیب انتگرال روی مسیر $C_+$ به روش جزء به جزء به صورت زیر محاسبه خواهد شد که جواب آن البته صفر می‌شود:

$$I = \omega \ln\left(1 - \left(\frac{\varepsilon q - \kappa}{\varepsilon q + \kappa}\right)^2 e^{-2qa}\right)\bigg|_{C_+}$$

$$- \int_{C_+} d\omega \ln\left(1 - \left(\frac{\varepsilon q - \kappa}{\varepsilon q + \kappa}\right)^2 e^{-2qa}\right)$$

$$= \lim_{r \to \infty} r \ln\left(1 - \left(\frac{\varepsilon q - \kappa}{\varepsilon q + \kappa}\right)^2 e^{-2qa}\right)$$

$$- \lim_{r \to \infty} \int_{-\frac{\pi}{2}}^{\frac{\pi}{2}} r \ln\left(1 - \left(\frac{\varepsilon q - \kappa}{\varepsilon q + \kappa}\right)^2 e^{-2qa}\right) d\theta$$

$$= 0 \qquad \qquad \text{۲۵}$$

اکنون جمله‌ی باقی‌مانده از معادله‌ی (۲۳) را با استفاده از تغییر متغیر مناسب $\xi = i\omega$، به صورت

$$E_{\text{Cas}}^{\text{TM}}(a) = -\frac{\hbar S}{8\pi^2} \int_0^\infty \kappa_\perp d\kappa_\perp$$

$$\times \int_{-\infty}^{+\infty} \xi d\ln\left(1 - \left(\frac{\varepsilon q - \kappa}{\varepsilon q + \kappa}\right)^2 e^{-2qa}\right) \qquad \text{۲۶}$$

می‌نویسیم. معادله‌ی (۲۶) را نیز با انتگرال‌گیری جزء به جزء به صورت زیر محاسبه می‌کنیم:

$$E_{\text{Cas}}^{\text{TM}}(a) = \frac{-\hbar S}{8\pi^2} \int_0^\infty \kappa_\perp d\kappa_\perp$$

$$\times \left[ \xi \ln\left(1 - \left(\frac{\varepsilon q - \kappa}{\varepsilon q + \kappa}\right)^2 e^{-2qa}\right)\bigg|_{\xi=-\infty}^{\xi=+\infty} \right.$$

$$\left. - \int_{-\infty}^{+\infty} \ln\left(1 - \left(\frac{\varepsilon q - \kappa}{\varepsilon q + \kappa}\right)^2 e^{-2qa}\right) d\xi \right]$$

$$= \frac{\hbar S}{4\pi^2} \int_0^\infty \kappa_\perp d\kappa_\perp \int_0^{+\infty} d\xi \ln\left(1 - \left(\frac{\varepsilon q - \kappa}{\varepsilon q + \kappa}\right)^2 e^{-2qa}\right), \qquad \text{۲۷}$$

که در آن از زوج بودن انتگرال‌ده با توجه به رابطه (۱۷) استفاده نموده‌ایم. با در نظر گرفتن مُد TE انرژی کازیمیر به شکل

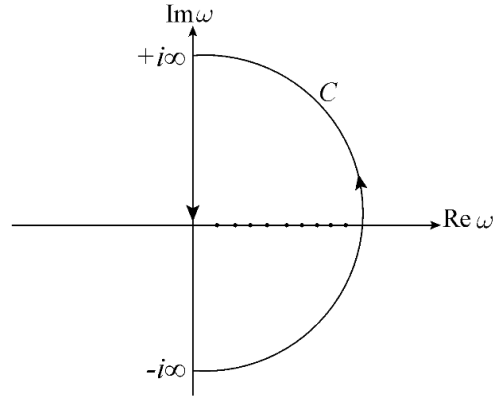

شکل ۳: مسیر پربندی $C$، که در آن نقاط بیان کننده‌ی ریشه‌های توابع مولد $\Delta^{\text{TM,TE}}(\omega_n, \kappa_\perp)$ می‌باشند.

که در خط دوم آن حد $b \to \infty$ را نیز گرفته‌ایم. برای حل انتگرال پربندی مختلط معادله‌ی (۲۲) مطابق شکل ۳، پربند $C$ همان طور که ذکر شد به مسیر $C_+$ و یک مسیر مستقیم روی محور موهومی شکسته می‌شود:

$$E_{\text{Cas}}^{\text{TM}}(a) = \frac{\hbar S}{8\pi^2 i} \int_0^\infty \kappa_\perp d\kappa_\perp$$

$$\times \left[ \int_{+i\infty}^{-i\infty} \omega d\ln\left(1 - \left(\frac{\varepsilon q - \kappa}{\varepsilon q + \kappa}\right)^2 e^{-2qa}\right) \right.$$

$$\left. + \underbrace{\int_{C_+} \omega d\ln\left(1 - \left(\frac{\varepsilon q - \kappa}{\varepsilon q + \kappa}\right)^2 e^{-2qa}\right)}_{I} \right]. \qquad \text{۲۳}$$

از آنجا که $C_+$ یک نیم دایره به شعاع $r \to \infty$ ($\omega = re^{i\theta}$) می‌باشد. از آنجا که اجسام نسبت به بسامدهای بالا (بالاتر از بسامد پلاسمای جسم) بطور شفاف رفتار می‌کنند در این حد خواهیم داشت

$$r \to \infty \qquad \qquad \text{۲۴}$$

$$\Rightarrow \varepsilon(\omega) \to 1, \ \kappa = q \to \infty, \ e^{-2qa} \to 0.$$

$$E_{\text{Cas}}(a) = \frac{\hbar S}{4\pi^2} \int_0^\infty \kappa_\perp d\kappa_\perp \int_0^\infty d\xi$$
$$\times \left\{ \ln\left[1 - r_{\text{TM}}^2\left(i\xi, \kappa_\perp\right) e^{-2qa}\right] \right.$$
$$\left. + \ln\left[1 - r_{\text{TE}}^2\left(i\xi, \kappa_\perp\right) e^{-2qa}\right] \right\}, \quad ۲۸$$

خواهد شد، که در آن $r_{\text{TM}}$ و $r_{\text{TE}}$ به ترتیب ضرایب بازتاب مربوط به مُدهای مغناطیسی عرضی و الکتریکی عرضی می‌باشند و به صورت زیر نوشته می‌شوند:

$$r_{\text{TM}}(i\xi, \kappa_\perp) = \frac{\varepsilon(i\xi) q(i\xi, \kappa_\perp) - \kappa(i\xi, \kappa_\perp)}{\varepsilon(i\xi) q(i\xi, \kappa_\perp) + \kappa(i\xi, \kappa_\perp)}$$
$$r_{\text{TE}}(i\xi, \kappa_\perp) = \frac{q(i\xi, \kappa_\perp) - \kappa(i\xi, \kappa_\perp)}{q(i\xi, \kappa_\perp) + \kappa(i\xi, \kappa_\perp)}. \quad ۲۹$$

اگر از رابطه‌ی فوق طبق معادله‌ی (۶) مشتق بگیریم حاصل، نیروی کازیمیر بین دو صفحه‌ی دی الکتریک در خلأ می‌باشد،

$$\frac{F_{\text{Cas}}(a)}{S} = \frac{\hbar}{2\pi^2} \int_0^\infty \kappa_\perp d\kappa_\perp \int_0^\infty d\xi, \quad ۳۰$$
$$\times q \left[ \frac{1}{1 - r_{\text{TM}}^{-2}(i\xi, \kappa_\perp) e^{2qa}} \right.$$
$$\left. + \frac{1}{1 - r_{\text{TE}}^{-2}(i\xi, \kappa_\perp) e^{2qa}} \right]$$

این رابطه همان فرمول معروف لیفشیتز است که ما به صورت سرراست و خودسازگار، بدون در نظر گرفتن فرض‌های پیچیده‌ی پراکندگی به دست آوردیم.

## نتیجه‌گیری

در این مقاله ما توانستیم فرمول معروف لیفشیتز، مربوط به انرژی کازیمیر بین دو صفحه‌ی دی الکتریک در دمای صفر مطلق، را با استفاده از تعریف صحیح و دقیق انرژی کازیمیر محاسبه نماییم. جزئیات استخراج این رابطه در مقاله‌ی اصلی لیفشیتز نیامده است. حذف بی‌نهایت‌ها و استخراج جواب فیزیکی متناهی در این مسأله پیچیدگی‌های خاصی دارد. مراجع متعددی که برای به دست آوردن جواب تلاش نموده‌اند، تا جایی که ما اطلاع پیدا کردیم، یا از تعریف ناصحیح انرژی کازیمیر استفاده کرده‌اند و این اشتباه با بی دقتی‌های دیگر پوشیده شده و یا روش‌های پیچیده و البته قابل تأمل پراکندگی استفاده نموده‌اند. در اکثر این موارد برخی بی‌نهایت‌ها به صورت دستی از مسأله خارج می‌شود. در این مقاله ما فرمول لیفشیتز را با استفاده از بازبهنجارش جعبه به صورت کاملاً سیستماتیک و خودسازگار به دست آورده‌ایم. در این روش از تعریف درست انرژی کازیمیر استفاده شده و همه‌ی بی‌نهایت‌ها (اعم از بی‌نهایت‌های وابسته به فاصله‌ی دو صفحه و بی‌نهایت‌های مستقل از آن) در خلال محاسبه و نه به صورت دستی حذف می‌شوند. فرمول (۳۰)، فشار وارد بر صفحات، همان رابطه‌ی لیفشیتز می‌باشد.

## تشکر و قدردانی

## پیوست:

در این پیوست، با استفاده از معادلات ماکسول و اعمال شرایط مرزی مناسب توابع مولد مُدها، $\Delta^{\text{TM}}$ و $\Delta^{\text{TE}}$، را محاسبه خواهیم نمود. بدین منظور میدان‌های الکتریکی و مغناطیسی را به شکل زیر در نظر می‌گیریم:

پ.۱
$$\mathbf{E}(t,\mathbf{r}) = \mathbf{E}(\mathbf{r})e^{-i\omega t}$$
$$\mathbf{B}(t,\mathbf{r}) = \mathbf{B}(\mathbf{r})e^{-i\omega t}.$$

روابط فوق را در معادلات ماکسول فضای آزاد جایگذاری می‌کنیم، که حاصل به شکل

پ.۲
$$\nabla . \mathbf{E}(\mathbf{r}) = 0$$
$$\nabla . \mathbf{B}(\mathbf{r}) = 0$$
$$\nabla \times \mathbf{E}(\mathbf{r}) - \frac{i\omega}{c}\frac{\partial \mathbf{B}(\mathbf{r})}{\partial t} = 0$$
$$\nabla \times \mathbf{B}(\mathbf{r}) + \frac{i\varepsilon(\omega)\omega}{c}\frac{\partial \mathbf{E}(\mathbf{r})}{\partial t} = 0,$$

خواهد بود. با مشتق‌گیری از دو معادله‌ی اخیر روابط زیر به دست می‌آید

پ.۳
$$\nabla^2 \mathbf{E}(\mathbf{r}) + \varepsilon(\omega)\frac{\omega^2}{c^2}\mathbf{E}(\mathbf{r}) = 0$$
$$\nabla^2 \mathbf{B}(\mathbf{r}) + \varepsilon(\omega)\frac{\omega^2}{c^2}\mathbf{B}(\mathbf{r}) = 0,$$

که دسته جوابهای متعامد کاملی به صورت

پ.۴
$$\mathbf{E}_s(\mathbf{r},\boldsymbol{\kappa}_\perp,\omega) = \mathbf{e}_s(z,\boldsymbol{\kappa}_\perp)e^{i\boldsymbol{\kappa}_\perp . \mathbf{r}_\perp}$$
$$\mathbf{B}_s(\mathbf{r},\boldsymbol{\kappa}_\perp,\omega) = \mathbf{b}_s(z,\boldsymbol{\kappa}_\perp)e^{i\boldsymbol{\kappa}_\perp . \mathbf{r}_\perp}$$

برای آنها (متناسب با مسأله‌ی مورد نظر ما) متصور است. با جایگذاری جوابهای (پ.۴) در معادله‌ی (پ..۳)، معادلات نوسانگر برای توابع $e_p, b_p$ به صورت

پ.۵
$$\mathbf{e}''_s(z,\boldsymbol{\kappa}_\perp) - \kappa^2 \mathbf{e}_s(z,\boldsymbol{\kappa}_\perp) = 0$$
$$\mathbf{b}''_s(z,\boldsymbol{\kappa}_\perp) - \kappa^2 \mathbf{b}_s(z,\boldsymbol{\kappa}_\perp) = 0$$

به دست خواهد آمد. علامت پرایم در عبارت بالا نشان دهنده‌ی مشتق‌گیری نسبت به $z$ می‌باشد. در ضمن در فضای بین دوصفحه $\varepsilon(\omega) = 1$ در نظر گرفته می‌شود. به همین ترتیب برای دو معادله‌ی دیگر ماکسول $\nabla . \mathbf{E}(\mathbf{r}) = 0$ و $\nabla . \mathbf{B}(\mathbf{r}) = 0$ خواهیم داشت

پ.۶
$$\mathbf{e}'_{s,z}(z,\boldsymbol{\kappa}_\perp) + i\kappa_x . \mathbf{e}_{s,x}(z,\boldsymbol{\kappa}_\perp)$$
$$+ i\kappa_y . \mathbf{e}_{s,y}(z,\boldsymbol{\kappa}_\perp) = 0$$
$$\mathbf{b}'_{s,z}(z,\boldsymbol{\kappa}_\perp) + i\kappa_x . \mathbf{b}_{s,x}(z,\boldsymbol{\kappa}_\perp)$$
$$+ i\kappa_y . \mathbf{b}_{s,y}(z,\boldsymbol{\kappa}_\perp) = 0.$$

اگر صفحه دی‌الکتریک ضخیم را ناحیه‌ی ۱ و فضای بین صفحات را ناحیه‌ی ۲ بگیریم، شرایط مرزی زیر را خواهیم داشت:

پ.۷
$$\mathbf{E}_{1t}(t,\mathbf{r}) = \mathbf{E}_{2t}(t,\mathbf{r}),$$
$$\mathbf{D}_{1n}(t,\mathbf{r}) = \mathbf{D}_{2n}(t,\mathbf{r}),$$
$$\mathbf{B}_{1t}(t,\mathbf{r}) = \mathbf{B}_{2t}(t,\mathbf{r}),$$
$$\mathbf{B}_{1n}(t,\mathbf{r}) = \mathbf{B}_{2n}(t,\mathbf{r}).$$

اولین شرط مرزی معادله‌ی (پ.۷) به پیوستگی $e_{s,x}, e_{s,y}$ در عبور از صفحات مرزی $z = \pm\frac{a}{2}$ می‌انجامد، که از معادله‌ی (پ.۶) معادل پیوستگی $e'_{s,z}$ است. دومین شرط مرزی، $\mathbf{D}_{1n}(t,\mathbf{r}) = \mathbf{D}_{2n}(t,\mathbf{r})$، باعث پیوستگی $\varepsilon e_{p,z}$ در عبور از صفحات می‌شود. از آنجایی که مولفه‌ی $z$ میدان الکتریکی تنها برای قطبش TM غیر صفر است (مطابق شکل ٤)،

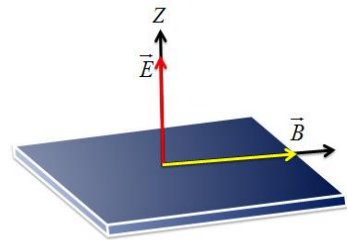

شکل ٤: قطبش مغناطیسی عرضی.

تنها از حالت مغناطیسی عرضی استفاده می‌کنیم و $e_{\text{TM},z}$ را بصورت زیر نمایش می‌دهیم:

$$\mathbf{e}_{\text{TM},z} = \begin{cases} Ae^{\kappa z} & z < \dfrac{-a}{2} \\ Be^{qz} + Ce^{-qz} & \dfrac{-a}{2} \leq z \leq \dfrac{+a}{2} \\ De^{-\kappa z} & z > \dfrac{+a}{2}. \end{cases} \quad \text{پ.٨}$$

با توجه به پیوستگی‌های که پیشتر توضیح داده شد، با اعمال شرایط مرزی در $z = \pm\dfrac{a}{2}$ به معادلات زیر می‌رسیم:

$$A\kappa e^{-\kappa\frac{a}{2}} = Bqe^{-q\frac{a}{2}} - Cqe^{q\frac{a}{2}}$$
$$-D\kappa e^{-\kappa\frac{a}{2}} = Bqe^{q\frac{a}{2}} - Cqe^{-q\frac{a}{2}}$$
$$A\varepsilon e^{-\kappa\frac{a}{2}} = Be^{-q\frac{a}{2}} + Ce^{q\frac{a}{2}} \qquad \text{پ.٩}$$
$$D\varepsilon e^{-\kappa\frac{a}{2}} = Be^{q\frac{a}{2}} + Cqe^{-q\frac{a}{2}}$$

که در آن

$$q^2 = \kappa^2_\perp - \frac{\omega^2}{c^2} \quad \text{و} \quad \kappa^2 = \kappa^2_\perp - \varepsilon(\omega)\frac{\omega^2}{c^2}$$
پ.١٠

شرط وجود جواب برای دسته معادلات (پ.٩) این است که دترمینان ضرایب صفر شود.، یعنی

$$\Delta^{\text{TM}}(\omega, \kappa_\perp, a) \equiv e^{-\kappa a} \qquad \text{پ.١١}$$
$$\times\left[(\varepsilon q + \kappa)^2 e^{qa} - (\varepsilon q - \kappa)^2 e^{-qa}\right] = 0$$

به همین ترتیب برای مُد TE نیز خواهیم داشت:

$$\Delta^{\text{TE}}(\omega, \kappa_\perp, a) \equiv e^{-\kappa a} \qquad \text{پ.١٢}$$
$$\times\left[(q + \kappa)^2 e^{qa} - (q - \kappa)^2 e^{-qa}\right] = 0.$$

[1–34]